# Thermal transport across grain boundaries in polycrystalline silicene: a multiscale modeling

Maryam Khalkhali[1], Ali Rajabpour[2], Farhad Khoeini[1]*

[1.] Department of Physics, University of Zanjan, Zanjan 45195-313, Iran

[2.] Mechanical Engineering Department, Imam Khomeini International University, Qazvin 34148–96818, Iran

**Abstract**

During the fabrication process of large scale silicene through common chemical vapor deposition (CVD) technique, polycrystalline films are quite likely to be produced, and the existence of Kapitza thermal resistance along grain boundaries could result in substantial changes of their thermal properties. In the present study, the thermal transport along polycrystalline silicene was evaluated by performing a multiscale method. Non-equilibrium molecular dynamics simulations (NEMD) was carried out to assess the interfacial thermal resistance of various constructed grain boundaries in silicene as well as to examine the effects of tensile strain and the mean temperature on the interfacial thermal resistance. In the following stage, the effective thermal conductivity of polycrystalline silicene was investigated considering the effects of grain size and tensile strain. Our results indicate that the average values of Kapitza conductance at grain boundaries at room temperature were estimated nearly $2.56\times10^9$ W/m$^2$K and $2.46\times10^9$ W/m$^2$K through utilizing Tersoff and Stillinger-Weber interatomic potentials, respectively. Also, in spite of the mean temperature whose increment does not change Kapitza resistance, the interfacial thermal resistance can be controlled by applying strain. Furthermore, it was found that, by tuning the grain size of polycrystalline silicene, its thermal conductivity can be modulated up to one order of magnitude.

**Keywords:** Polycrystalline silicene, Effective thermal conductivity, Non-equilibrium molecular dynamics, Grain boundary, Grain size, Strain.

## 1. Introduction

Nowadays, novel two-dimensional materials have attracted widespread research interest due to their promising application potential in nanotechnology [1-4]. The successful exfoliation of graphene [5, 6] as a one-atom-thick planar sheet with fascinating properties [7-9] served as a milestone in experimental attempts to synthesize the new class of 2D materials [10-13].

Silicene, a honeycomb structure of silicon elements, has emerged as a favorable monolayer material because of its outstanding physical and chemical properties [3, 14, 15] as well as its compatibility with current silicon-based electronics [16]. Although theoretical possibility of silicene was predicted by Takeda et al. [17] in 1994 but it was successfully grown on various metal substrates in the recent decade [3, 18-20].

Owing to the structural similarity of silicene and graphene, silicene can be a satisfactory alternative to graphene. Furthermore, the slightly buckled structure of silicene and the existence of mixed $sp^2$-$sp^3$ hybridization lead to some new features [21]. For instance, despite graphene, the band gaps of silicene can be opened and tuned when exposed to an external electric field, which propose it as desirable building-blocks, such as field effect transistors [22, 23], solar cells [24], energy storage [25], reusable molecule sensors [26], and Li-ion batteries [27]. In addition, the physics in quantum phase transition can be explored with the interaction between the electromagnetic field and spin–orbit coupling in silicene [28, 29]. Moreover, this slightly buckled structure will significantly alter the thermal conductivity of silicene due to the breaking the symmetry of the out-of-plane direction. Thus, unlike graphene, silicene exhibits a low thermal conductivity [30]. Because the effective procedure to improve the thermoelectric performance is to reduce the thermal conductivity with maintaining the electronic transport features, silicene may show a great advantage as a promising thermoelectric material [30, 31].

It is notable that the reported distinguished properties of silicene belong to single crystal and defect-free samples, while various types of defects are formed typically during the fabrication process which may considerably affect electrical, mechanical, thermal, and other properties of silicene [32-34].

Among all the developed fabrication methods, the chemical vapor deposition (CVD) is a common approach to synthesize different types of two-dimensional atomic crystals due to its simplicity, the potential to be applied in large-scale and high-quality production and its rather low expenses [35,36]. It is worth mentioning that CVD proses inevitably lead to the formation of polycrystalline structures. In the polycrystalline morphology, grain boundaries appear where grains with different crystalline orientations face each other. The grain boundaries expand across the structure because their more distinct lattices than that of pristine grains are regarded as topological defects. These topological defects can scatter the phonons and can also drastically change the thermal conduction properties of product [37] or alter the mechanical [38,39], and electronic features of the constructed structure.

Mortazavi et al. [40] investigated the thermal transport of polycrystalline $MoS_2$ by developing a multiscale method. They first performed the molecular dynamics simulations in order to explore Kapitza thermal conductance of different types of grain boundaries which are detectable in CVD constructed $MoS_2$. Next, in order to study the effective thermal conductivity of polycrystalline samples at macroscopic level, they designed continuum models of $MoS_2$ films utilizing the finite element method. Consequently, they observed that thermal conductivity of samples can be modulated by changing their grain size. Bazrafshan et al. [41] developed a combined NEMD atomistic -continuum multiscale modeling to engineering of thermal transport in polycrystalline graphene. They also examine the impact of nitrogen and boron doping, grain size, and mechanical strain on the effective thermal conductivity of polycrystalline graphene films. Their results indicated that Kapitza conductance and the thermal conductivity of polycrystalline graphene with nano-sized grains is not affected with nitrogen and boron doping. Also they represented that the interfacial thermal resistance has the significant role in thermal transport within polycrystalline graphene with small grain sizes. Gao et al. [35] carried out Green−Kubo equilibrium molecular dynamics simulations to measure the thermal conductivity of polycrystalline and amorphous silicene. According to their report, polycrystalline silicene demonstrates extremely low thermal conductivity compared with amorphous silicene and other polycrystalline silicon nanomaterials with the identical grain size. Bagri et al. [42] utilized non-equilibrium molecular dynamics simulations to evaluate the effect of tilt grain boundaries on the thermal conductivity of graphene. They observed a temperature jump at the grain boundary in the temperature profile and showed that the thermal conductivity of graphene slightly hinges on the orientation of the grains. Sledzinska et al. [43] reported a successful growth of polycrystalline $MoS_2$ with tunable morphologies. They managed the thermal conductivity of the sample as a function of grain orientation and recorded low thermal conductivity of 0.27 W/mK in polycrystalline $MoS_2$. Roy et al. [44] used the non-equilibrium molecular dynamics simulation to study the role of grain size in the thermal conductivity of polycrystalline Silicene. Their findings revealed that the thermal conductivity of the polycrystalline silicene with ultra-fine nano-grained is more sensitive to the grain size compared with the sample with larger grain size. Therefore, with increasing the grain size, the thermal conductivity increases and finally converges to a certain value. Li et al. [45] investigated the thermal and mechanical properties of grain boundary in the hybrid interface between graphene and hexagonal boron-nitride with utilizing multiscale approaches consisted of molecular dynamics simulation, density functional theory and classical disclination theory. They concluded that the graphene and hexagonal boron-nitride grains matched by non-symmetric grain boundary which is composed of periodic pentagon-heptagon defects. Also they found that the mismatch angle of grains and the direction of thermal flux affect the thermal transfer efficiency. Cui et al. [46] employed molecular dynamics simulations to study the thermal properties of Silicene nanomesh. According to their results, the thermal conductivity of silicene nanomesh is much lower in comparison with pristine silicene. Besides, they showed that the thermal conductivity of the silicene nanomesh decreases with the increment of porosity. Chen et al. [47] examined the effect of uniaxial cross-plane strain on the Kapitza thermal resistance of few-layer graphene. Their calculations indicated that Kapitza resistance increases by applying the strain, from compressive to tensile. Jhon et al. [48] applied non-equilibrium molecular dynamics simulations to probe the impact of temperature on the thermal boundary conductance of polycrystalline graphene for different types of grain boundary and various grain sizes. The results of their study exhibits that Kapitza conductance enhances by increasing the temperature.

Among the diverse properties of silicene, its thermal property plays a crucial role in the thermal management of silicene-based nano-devices, not only in their efficiency but also for the safety. Nevertheless, despite the intense study on the thermal properties of silicene sheet and silicene nanoribbons, some limited research has been devoted to the thermal properties of polycrystalline silicene and the effect of the interfacial thermal resistance (Kapitza thermal resistance) of the grain boundaries on the thermal conductivity of polycrystalline silicene so far.

In the present study, we investigate the thermal properties of polycrystalline silicene through performing a multiscale method, consisting of non-equilibrium molecular dynamics simulations (NEMD) and solving continuum heat conduction equation. We first applied classical NEMD simulations in order to compute the interfacial thermal resistance (Kapitza thermal resistance) of six different constructed grain boundaries in silicene. Following that, we intensively explored the effect of tensile strain and the mean temperature on the interfacial thermal resistance. Finally, we made a continuum model of polycrystalline silicene based on the MD provided data and examine the effective thermal conductivity of the model by taking into consideration the effects of grain size and mechanical strain.

## 2. Simulation method

Based on first-principles calculations [34, 49-51] and experimental evidence via high-resolution electron microscopy [52, 53], various types of dislocation cores may exist in the grain boundaries of the two-dimensional polycrystalline materials. It is worth noting that the existence of a wide variety of dislocation cores depends on grains orientation as well as distances between the atoms in the two sides of the grain boundary. Among these diverse configurations, we have explored a common pentagon-heptagon defect pairs with different defect concentrations along the grain boundaries. Fig. 1 illustrates the atomic structures of six different grain boundaries consisted of pentagon-heptagon defect pairs. Also, we have evaluated both the symmetrical and asymmetrical grain boundaries for pentagon-heptagon defect pairs. As depicted in Fig. 1 for both the symmetrical and the asymmetrical grain boundaries, the defect concentration along the grain boundaries gradually increases. There are two hexagonal rings, which separate the two 5-7 defect pairs for the least defective ones, and for the most defective cases, there exist no hexagonal rings, which separate the two 5-7 dislocation cores. Also all constructed structures are periodic along the grain boundary direction.

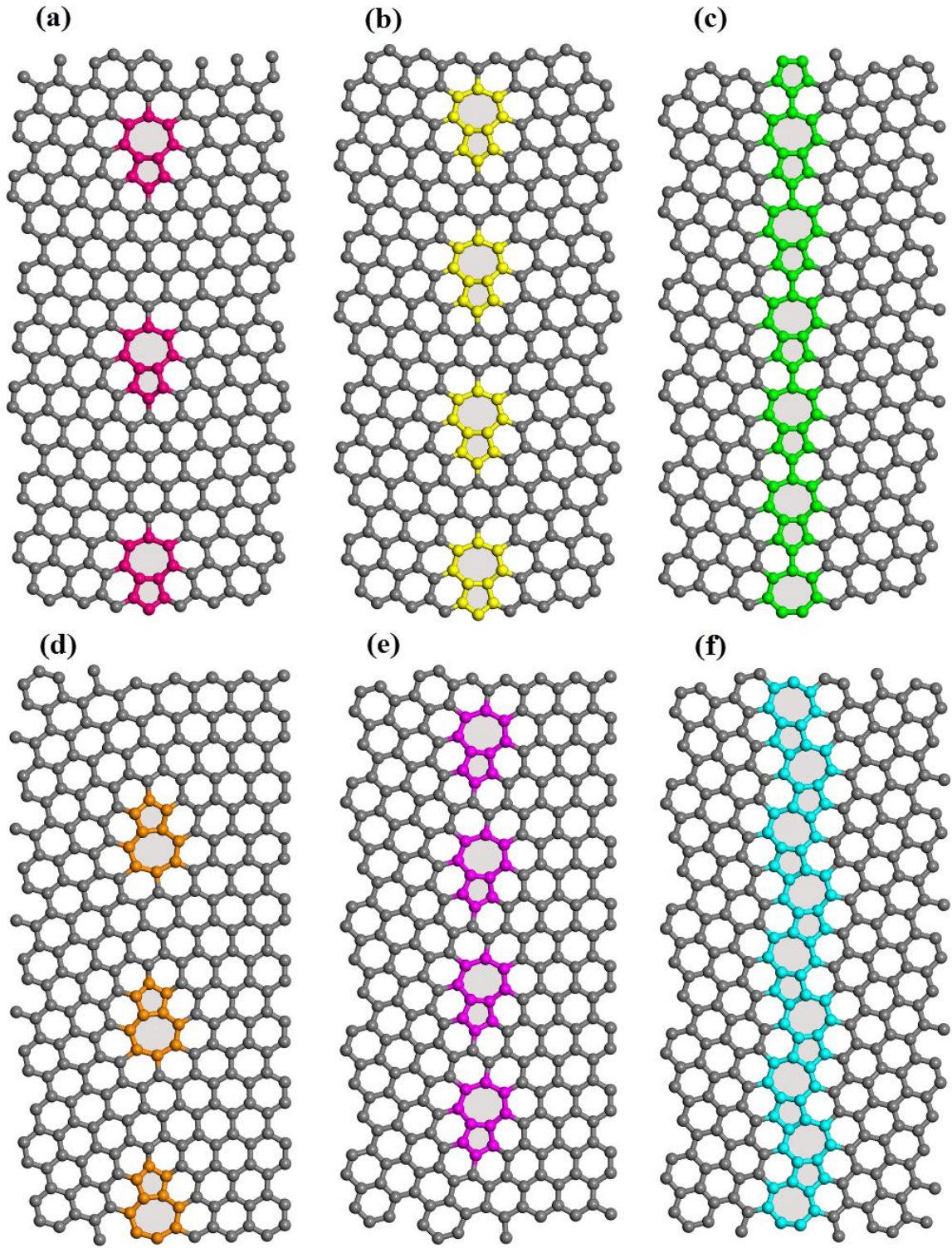

Fig. 1. Atomic structures of six different grain boundaries consist of pentagon-heptagon defect pairs with different defect concentrations: (a) 5-7-6-6-s, (b) 5-7-6-s, (c) 5-7-5-7-s, (d) 5-7-6-6-a, (e) 5-7-6-a, and (f) 5-7-5-7-a.

In this research, molecular dynamics simulation is performed using the Large Scale Atomic/Molecular Massively Parallel Simulator (LAMMPS) package [54] in order to calculate the interfacial thermal resistance of various grain boundaries in silicene. In the MD method, employing the appropriate potential functions is required for the accuracy of predictions. Previous studies have shown that, Stillinger-Weber and Tersoff potentials have been widely utilized to evaluate the thermal properties of the silicon-based materials. Unfortunately, in the case of silicene, the commonly used Stillinger-Weber potential cannot maintain the structure of the silicene. Therefore, it is not proper for investigating the thermal conductivity of monolayer silicene. As a result, we have employed optimized Stillinger-Weber potential which is recently reparametrized by Zhang et al. [31] which can precisely reproduce the buckled structure of silicene and the phonon dispersion computed from ab initio. On the other hand, to evaluate the sensitivity of the results to the chosen interaction potential, we have employed Tersoff [55] potential as another potential to compare the results. In addition, Newton's equations of motion are integrated via the velocity Verlet algorithm [56] with a time step of 0.5 fs for Stillinger-Weber potential and 1 fs in the case of Tersoff potential. Also, the periodic boundary condition is employed in the X direction and the free boundary conditions are set in Y and Z directions.

In order to rearrange the atoms at grain boundaries and get the stable structures, energy minimization performed and the systems annealed from 1 to 10 K for 150 ps in an NVE ensemble using Langevin thermostat. Then, the structures have been relaxed at room temperature (300 K) for 1 ns under NVT ensemble and coupling to Nose-Hover temperature thermostat. In order to generate the temperature gradient and consequently, non-equilibrium heat current across the system, the sample was partitioned into 35 slabs along Y direction and the atoms at the two ends were fixed. Adjacent to these fix regions, there exist hot and cold slabs which are coupled to Nose-Hoover thermostat using NVT ensemble to set the temperature at T+ΔT/2 (315 K) and T-ΔT/2 (285 K), respectively, also NVE ensemble is applied to the rest of the slabs. The mentioned condition has been applied to the system up to the steady-state regime is achieved and leads to a constant heat-flux. It is worth mentioning that the system has been simulated for the entire 10 ns and the first 4 ns have been discarded as pre-equilibration step. Molecular dynamics setup for calculating the boundary resistance of the grain boundary has been shown in Fig. 2.

The boundary resistance of the grain boundary can be calculated using the following equation [57, 58]:

$$R_K = \frac{\Delta T_{GB}}{J_Y}, \quad (1)$$

where $\Delta T_{GB}$ is the temperature drop at the interface which is the signature of the interfacial thermal resistance and can be obtained from the discontinuity of the 1D temperature profile along the sample length and $J_y$ is the heat flux along Y direction. In order to calculate the heat flux, the accumulative energy was added and subtracted from baths were recorded every 1000 timesteps and plotted versus time. The slopes of the linear fitting to energy diagrams are equal to the heat current. We assumed the thickness of 4.2 Å for single-layer silicene.

Finally, the acquired interfacial thermal resistance values were used in two models: 1D thermal resistance model and continuum modeling for the exploring of effective thermal conductivity of polycrystalline silicene sheet.

In the 1D thermal resistance model, the thermal resistance of polycrystalline silicene sheet can be considered as the sum of the thermal resistance of the grain and the interfacial thermal resistance along the grain boundary according to the following equation [41]:

$$\frac{n \times GS}{\kappa_{eff}} = n \frac{GS}{\kappa_G} + (n-1) \frac{1}{\kappa_{GB}}, \quad (2)$$

where n is the average number of grains along the Y direction, $K_G$ is the thermal conductivity of the grain, $K_{GB}$ is the Kapitza thermal conductance and GS is grain size which is defined as $GS = \sqrt{A/N}$ (A is the total area of the silicene sheet and N is the number of grains in polycrystalline silicene sample).

When the number of grains is very large ($n \to \infty$), eq. 2 can be simplified and effective thermal conductivity of polycrystalline silicene sheet can be calculated by [40]:

$$\kappa_{eff} = \frac{\kappa_G \times \kappa_{GB} \times GS}{\kappa_G + \kappa_{GB} \times GS}, \quad (3)$$

where an equal thermal conductivity of 41 W/mK (the one of pristine silicene sheet) [59] was assigned for all grains and the average value of the Kapitza thermal conductance for the six constructed grain boundary -which are computed through the MD- is considered as the thermal conductance of the grain boundaries.

In the 2D heat conduction continuum model, we have utilized Voronoi algorithm to produce polycrystalline silisene. For the evaluation of heat transfer, 2D heat conduction equation in steady-state, $\nabla^2 T = 0$, within each grain has been considered. As already discussed, the thermal conductivity of 41 W/mK was assumed for all grains and to define the thermal conductance of the grain boundary, the NEMD results for six different grain boundaries averaged. Also in accordance with NEMD setup, temperatures at the left and right boundaries of polycrystalline silisene were set to $T_H = 315$ K and $T_C = 285$ K, respectively. By numerically solving heat conduction via finite volume method, effective thermal conductivity of polycrystalline silicene sheet has been obtained.

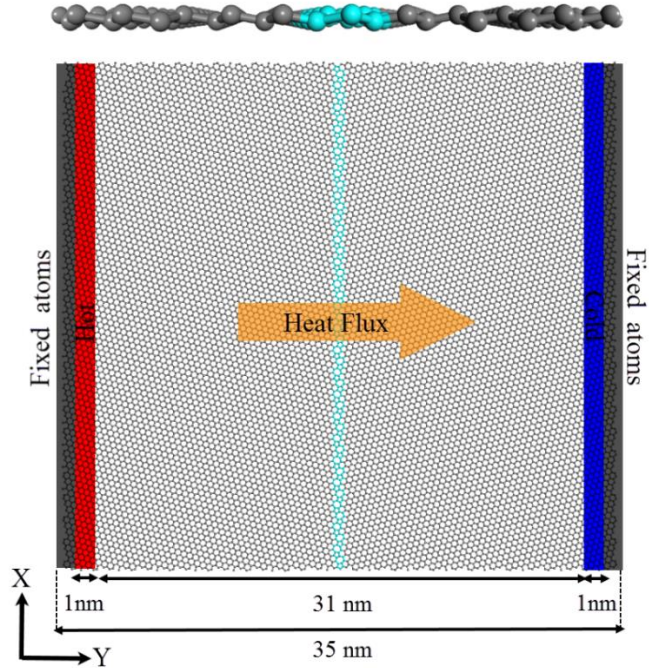

Fig. 2. Molecular dynamics setup for calculation of the boundary resistance of the grain boundary. Top panel: the side view of the system with 5-7-5-7-a defects. Bottom panel: the top view of the system with the same defects.

## 3. Results and discussions

We used a multiscale method consisting of non-equilibrium molecular dynamics simulations and solving continuum heat conduction equation in order to intensively investigate the effects of grain size and tensile strain on the thermal transport along the polycrystalline silicene.

The steady-state 1D temperature profiles of different constructed grain boundaries along Y direction have been presented in Fig. 3. A linear temperature gradient has been observed in the area away from the heat baths, and because of the phonon scattering with the heat baths, there exists nonlinearity near the two ends. Also, the most distinguished features of the all temperature profiles are the discontinuity at the middle of the sample, due to the existence of the grain boundary. The established temperature jump for all the samples are represented in Fig.3. It can be seen that for both the symmetric and non-symmetric grain boundaries, with increasing the defect concentration along the grain boundary, the temperature gap slightly increases. Furthermore, the temperature jump of the non-symmetric grain boundaries is explored to be higher than that of symmetric ones at the identical defect concentration.

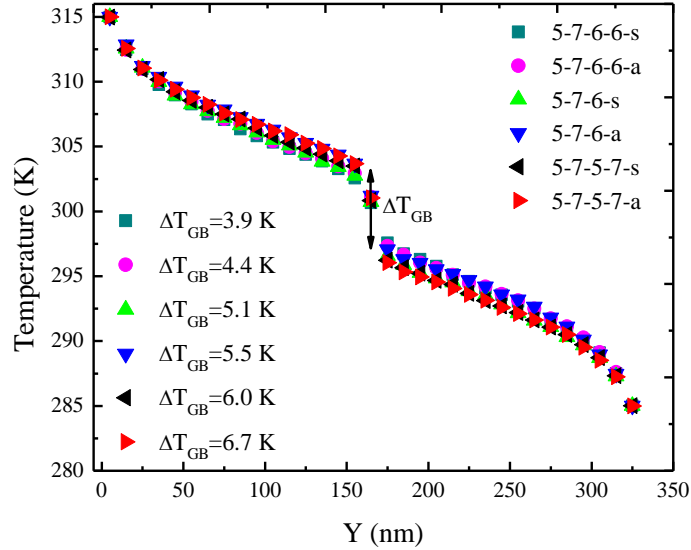

Fig. 3. The steady-state 1D temperature profiles of different constructed grain boundaries along Y direction with the same lengths of 35 nm at T=300 K and ΔT=30 K.

In Fig. 4, accumulative added energy to the hot segment and subtracted energy from the cold layer of the samples consist of non-symmetric grain boundaries with different defect concentration are illustrated. Besides, the heat current ($\frac{dE}{dt}$) is computed for each sample as the slope of the linear fit to energy profile, and depicted in Fig.4. As expected, by increasing the defect concentration, the magnitude of energy flux slightly decreases. As it is observed, the gradient of added energy with respect to time for each sample is equal to the rate of subtracted energy from that, witch confirms the conservation of energy in the NEMD simulations.

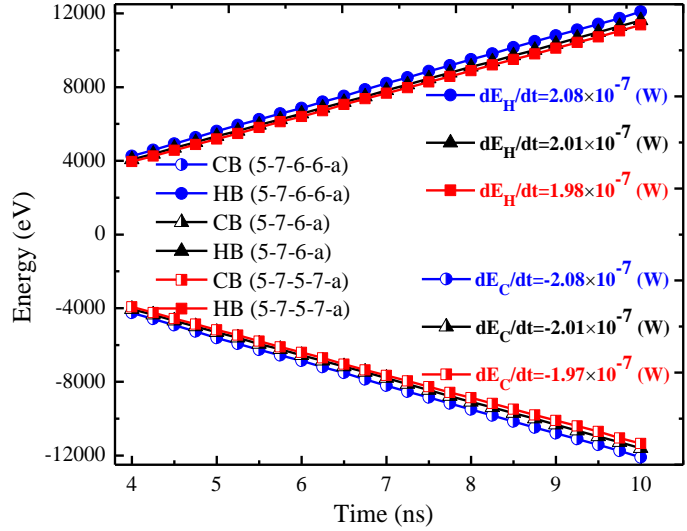

Fig. 4. Accumulative added energy to the hot layer and subtracted energy from cold layer as a function of simulation time of the silicene films consist of non-symmetric grain boundaries with different defect concentration at T=300 K and ΔT=30 K.

The interfacial thermal resistance of different constructed grain boundaries at room temperature is represented in Fig. 5. We have used both Stillinger-Weber [31] and Tersoff [55] potentials, to explore the sensitivity of the results to the chosen interaction potential. The values of the Kapitza thermal resistance obtained from Tersoff potential for symmetric grain boundaries varies from 28.5±0.3 ($\times 10^{-11}$ m$^2$K/W) to 43.8 ±1.6 ($\times 10^{-11}$ m$^2$K/W) at the least and most defect concentration, respectively. Furthermore, the amounts of Kapitza thermal resistance for non-symmetric grain boundaries ranges from 32.7±0.4 ($\times 10^{-11}$ m$^2$K/W) to 49.9±2.4 ($\times 10^{-11}$ m$^2$K/W) at the least and most defect concentration, respectively. As expected, by increasing the defect concentration along the grain boundary, for both the symmetric and non-symmetric grain boundaries, the thermal resistance increases. The previous change has happened because of the phonon-defect scattering. Also, it is found that, the thermal resistance of each non-symmetric grain boundaries is higher than the symmetric one at identical defect concentration. This higher interfacial thermal resistance is because of the higher phonon–phonon scattering through the non-symmetric grain boundaries, which arises from the difference of the phonon spectrum in two sides of the grain boundary. The findings are both reliable and in line with the results of Mortazavi et al. [40]. Furthermore, both Stillinger-Weber and Tersoff potentials have revealed nearly the same values for the interfacial thermal resistance of silicene grain boundaries at room temperature and the results are found to be insensitive to the chosen interaction potential.

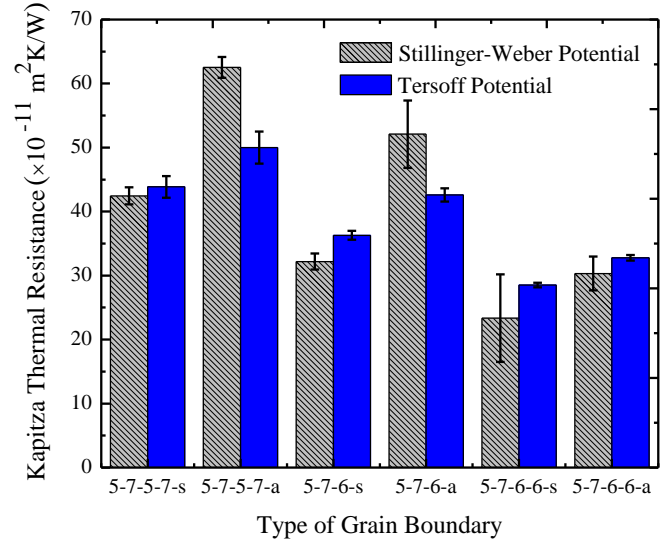

Fig. 5. The interfacial thermal resistance of different constructed grain boundaries at T=300 K and ΔT=30 K using Stillinger-Weber and Tersoff MD potentials.

In order to achieve a better elucidate of governed physical phenomena, we have calculated phonon density of states of two groups of atoms, belonging to two sides of the grain boundary. The phonon power spectral density has been obtained by computing the Fourier transform of autocorrelation function of the velocity of atoms corresponding to two sides of the grain boundary as follow [60, 61]:

$$P(\omega) = \sum_i \frac{m_i}{k_B T_{MD}} \int_0^\infty e^{-i\omega t} < v_i(t).v_i(0) > dt, \qquad (4)$$

where $\omega$ is the angular frequency, $m_i$ is the mass of atom i and $v_i$ is the velocity of the ith atom.

Figs. 6(a) and (b) show the phonon density of states of two sides of the most defected symmetric and non-symmetric samples (5-7-5-7-s and 5-7-5-7-a). As illustrated in figures, there exist mismatches between the two spectra of left and right sides of the grain boundary for each samples. Thus, it can be concluded that; the interfacial thermal resistance leads to the phonons scattering through the interface, which is in good agreement with previous results in the literature [62].

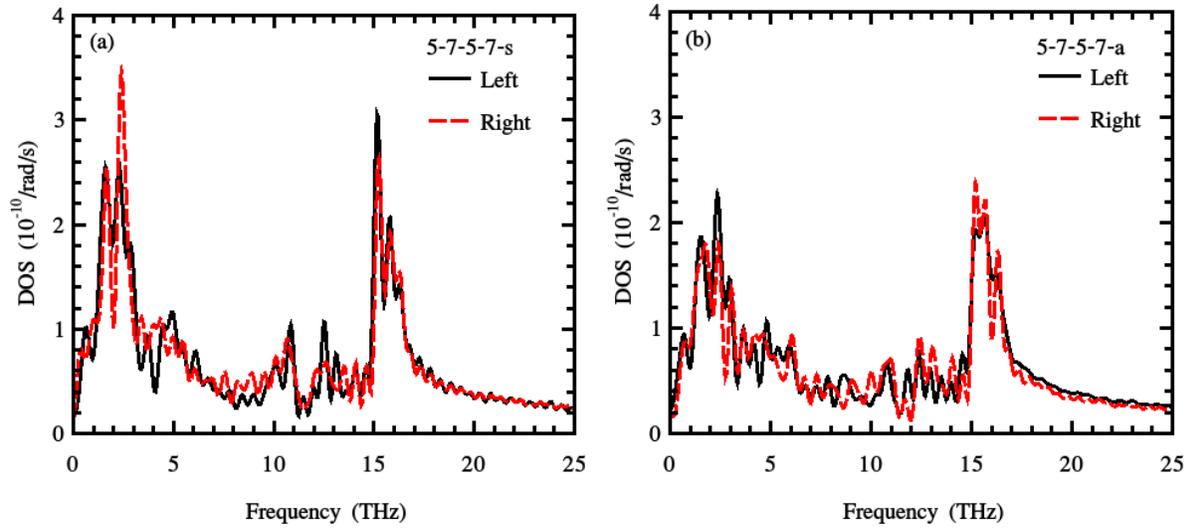

Fig. 6. Phonon power spectral density of states at two sides of 5-7-5-7-s (a) and 5-7-5-7-a (b) grain boundaries at T=300 K

In order to explore the sensitivity of the Kapitza thermal resistance to the mean temperature in silicene grain boundaries, we have raised the mean temperature from 300 to 700 K using the Stillinger-Weber potential. In Fig. 7, the NEMD predictions for the interfacial thermal resistance of various constructed grain boundaries as a function of mean temperature have been demonstrated. It is found that, by increasing the temperature up to 700 K, no significant alternation observed in the Kapitza thermal resistance of various silicene grain boundaries. The interfacial thermal resistance in all six types of grain boundaries depicted no changes and slight fluctuation was in the range of the statistical uncertainties of calculations. It is of note that Jhon et al. [48] reported that the boundary resistance in polycrystalline graphene decreases as the temperature increased but it is not found in the case of poly crystalline silicene, maybe it arose from the differences of silicene with graphene in terms of thermal properties. Also, as discussed before, the 5-7-5-7-a which is the most defective non-symmetric sample exhibits the highest interfacial thermal resistance and 5-7-6-6-s that is the least defective symmetric structure illustrates the lower Kapitza resistance.

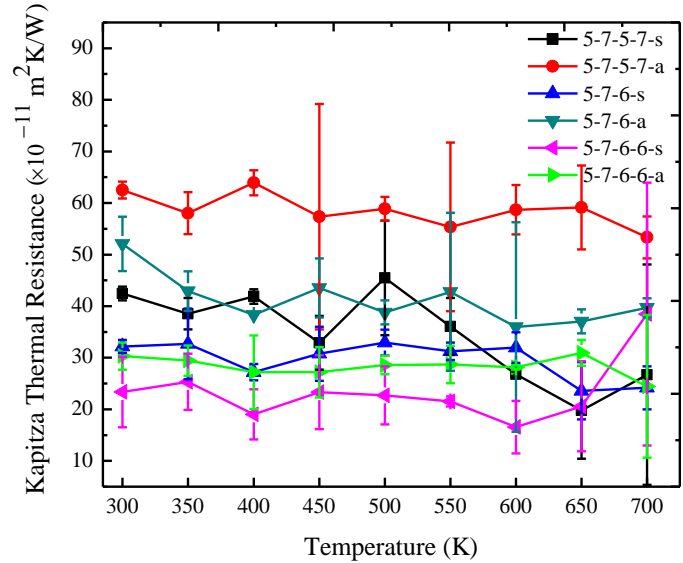

Fig. 7. The Kapitza thermal resistances of different constructed grain boundaries a function of mean temperature using the Stillinger-Weber potential; ΔT=30 K.

Studying the interfacial thermal resistance of silicene grain boundaries under an extreme condition such as strain is required for empirical applications. Therefore, in this step, we have investigated the effect of tensile strain on the Kapitza thermal resistance of the constructed grain boundaries. The strain is defined as [63]:

$$\varepsilon_y = \frac{dL}{L}, \quad (5)$$

where $L$ is the initial length of the silicene sheet and $dL$ is the change in length because of stretching one or both ends of the sample along the Y-direction. For imposing the tensile strain, the first segment of the silicene sheet is fixed and the last segment starts to stretch along the sheet length with the stretching velocity of 0.005 Å/ps.

In Fig. 8, the results of NEMD simulations for constructed grain boundaries under the tensile strain have been illustrated. It is of note that, also, we applied strain from 0.01 to 0.08 to each sample but critical value of tensile strain in some structures was less than 0.08. For instance, the 5-7-6-6-a structure stands tensile strain just up to 0.06. It is found that, the tensile strains applied parallel to the heat flux directions increases the Kapitza thermal resistance of the silicene grain boundaries and the minimum interfacial thermal resistance of the all samples is approximately found in strain-free constructions. Table 1 indicates the values of imposing tensile strain which causes the maximum increment in the Kapitza thermal resistance of the constructions.

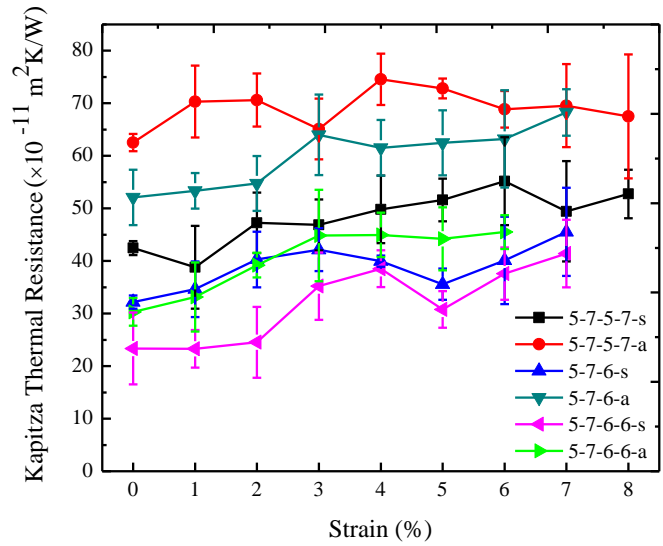

Fig. 8. The interfacial thermal resistances of six constructed silicene grain boundaries under tensile strain at T=300 K and ΔT=30 K

Table 1. Values of imposing tensile strain, which causes the maximum increment in the Kapitza thermal resistance of the samples.

| Grain Type | 5-7-5-7-s | 5-7-5-7-a | 5-7-6-s | 5-7-6-a | 5-7-6-6-s | 5-7-6-6-a |
|---|---|---|---|---|---|---|
| Strain | 0.06 | 0.04 | 0.07 | 0.07 | 0.07 | 0.06 |
| Increment of the Kapitza resistance (%) | 30 | 19 | 41 | 31 | 77 | 50 |

Fig. 9 represents the temperature profiles for two polycrystalline silicene sheet with grain sizes of GS =2 and 1000 nm, considering a grain thermal conductivity of $\kappa_G = 41$ W/mK and a Kapitza thermal conductance at grain boundaries of $2.46 \times 10^9$ W/m$^2$K. It is obvious that, for a sample with small grain size of 2 nm, the temperature distribution is approximately uniform inside each grain and a constant temperature can be assigned to individual grains. In contrast, for polycrystalline silicene film with large grain sizes, the temperature profile is not uniform inside each grain and a temperature gradient can be seen inside individual grains. These observations imply the fact that, at grain sizes lower than 10 nm the thermal resistance at grain boundaries dominates over the grains' thermal resistance but when the grain size increases, the effect of grain boundaries resistance is considerably weakened.

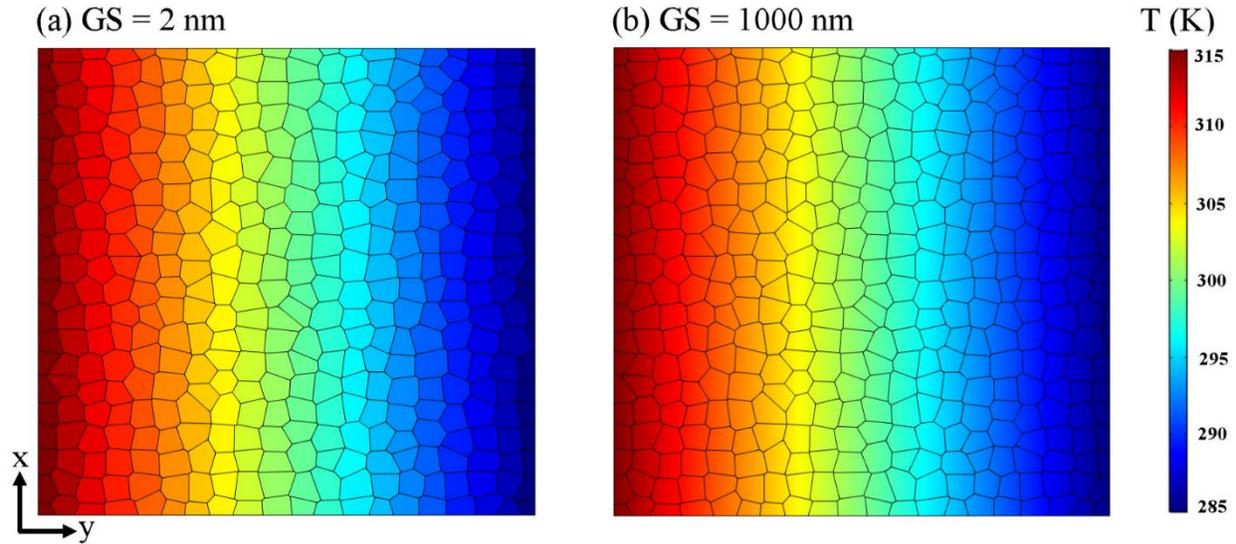

Fig 9. NEMD/2D continuum multiscale modeling results for the comparison of grain size effect on the temperature profile for polycrystalline silicene with grain sizes of 2 nm (a) and 1000 nm (b) at T=300 K and ΔT=30 K; $\kappa_G$ and $\kappa_{GB}$ were set to 41 W/mK and 2.46×10$^9$ W/m$^2$K, respectively.

In the present study, the thermal conductivity of polycrystalline silicene as a function of grain size has been evaluated by the 2D continuum multiscale model and 1D thermal resistance approach. In Fig. 10 the effective thermal conductivity of polycrystalline silicene with various grain sizes based on proposed models are represented. It can be observed that, the effective thermal conductivity of polycrystalline silicene exhibits increment with the increasing the sizes of grains. Furthermore, for samples with the grain sizes lower than 100 nm, the thermal conductivity increases drastically by increasing the grain size, but by increasing the grain size in the range of 100 to 1000 nm, the increase in the thermal conductivity declines and eventually achieves a plateau, near to the thermal conductivity of pristine silicene. As mentioned earlier, since we supposed identical thermal conductivity for all grains (41 W/mK) [59], the key factor which assign the effective thermal conductivity of polycrystalline film is thermal conductance of the grain boundaries. Besides, it is demonstrated that, when the grains sizes increase, the thermal conductivity of polycrystalline silicene changes nearly from 4.9 to 40.5 W/mK, specifying that by tuning the grain size of polycrystalline silicene, its thermal conductivity can be modulated up to one order of magnitude. Also, as represented in Fig. 10 the results based on 1D thermal resistance approach are in line with findings of 2D continuum multiscale model showing more suitability of the 1D thermal resistance model due to its simplicity. Our results are in line with refs. [40, 41].

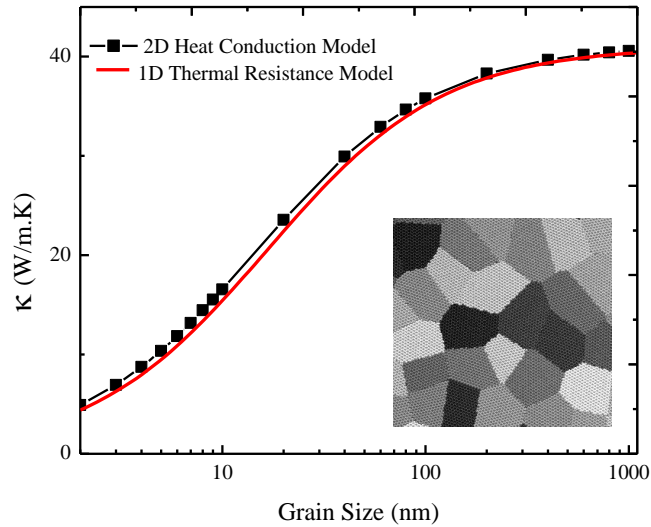

Fig.10. The effective thermal conductivity of polycrystalline silicene with various grain sizes based on 2D heat conduction and 1D thermal resistance models; grain sizes are set to 1, 2, 3, …, 10, 20, 40, …,100, 200, 400, …,1000 nm.

Eventually, we have examined the effect of tensile strain on the thermal conductivity of polycrystalline silicene. To this aim, we used a NEMD-2D continuum multiscale model. Fig. 11 depicts the thermal conductivity of polycrystalline silicene with different grain size of 2, 6, 10, 60 and 600 under tensile strain from 1% to 6%. It is of note that, for the grains, we have assumed the thermal conductivity of under strain pristine silicene [59] and for the thermal conductance of grain boundaries, we have averaged the NEMD results for the under strain constructed samples. As it is represented, when the grain size increases, polycrystalline silicene exhibits thermal response to tensile strain similar to pristine silicene, in a way that the thermal conductivity of polycrystalline silicene with grain size of 60 and 600 nm, increase with applying tensile strain up to 6%. This distinct response of silicene to tensile strain can be attributed to its buckled structure. When silicene started to stretch, at first its bucked structure started to become less buckled and in-plane stiffness of silicene increased, consequently the thermal conductivity enhanced. Following that, with more increase the tensile strains, the buckled structure of silicene would become flattened, thus the in-plane stiffness of silicene decreased and led to the decrement of the thermal conductivity.

It is obvious that, for the polycrystalline silicene with grain size smaller than 10 nm, by applying the strain up to 6% no remarkable changes observed on the effective thermal conductivity. As discussed earlier, the grain boundaries play the main role on the thermal conductivity of polycrystalline silicene with grain sizes smaller than 10 nm, and the dominant factor that defines the thermal conductivity is the thermal conductance of the grain boundaries [40, 41].

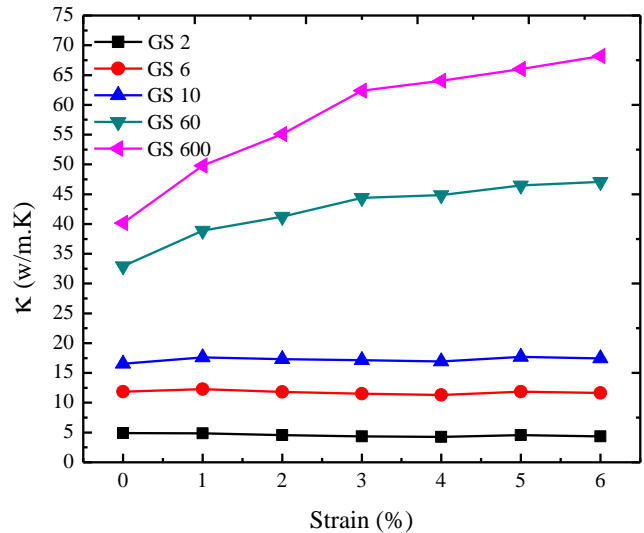

Fig.11. The thermal conductivity of polycrystalline silicene with different grain sizes of 2, 6, 10, 60 and 600 under tensile strain at T=300 K and ΔT=30 K.

**Conclusions**

In the present study, the thermal transport through polycrystalline silicene with performing a multiscale method consisting of classical NEMD and solving continuum heat conduction equation was intensively explored.

Extensive NEMD simulations were conducted so as to investigate the interfacial thermal resistance of six various constructed grain boundaries in polycrystalline silicene as well as to examine the effects of tensile strain and the mean temperature on the thermal resistance. The average values of Kapitza conductance at grain boundaries computed almost $2.56 \times 10^9$ W/m$^2$K and $2.46 \times 10^9$ W/m$^2$K by using Tersoff and Stillinger-Weber interatomic potentials, respectively. Furthermore, the results reveal that increasing the mean temperature of the sample does not affect the Kapitza resistance of the grain boundaries. In addition, it is found that, the interfacial thermal resistance of the grain boundaries can be tuned by applying tensile strain.

In order to examine the thermal properties of polycrystalline silicene film, the continuum model of polycrystalline silicene was constructed and its effective thermal conductivity was explored by taking the effects of grain size and tensile strain into consideration. It is worth mentioning that, the NEMD obtained results were used to determine the thermal conductance of grain boundaries of polycrystalline sample. The results revealed that, the thermal conductivity of polycrystalline silicene changes from 4.9 to 40.5 W/mK for the grains sizes of 2 to 1000 nm. Consequently, the effective thermal conductivity of polycrystalline silicene is adjustable by one order of magnitude with tuning its grain size. Also, the acquired results of the 2D continuum multiscale model were compared with those acquired from 1D thermal resistance approach, through which an acceptable agreement was observed.

Moreover, the thermal conductivity of polycrystalline silicene with large grain size (60 and 600 nm) increased after applying tensile strain up to 6%. For the sample with small grain size (less than 10 nm), the strain did not affect the thermal conductivity because the thermal conductance of grain boundaries plays a crucial role in the thermal transport properties of polycrystalline samples.

*E-mail: khoeini@znu.ac.ir